\documentstyle[12pt,a4]{article}
\title{\bf Relativistic Quantum States of an Electron with Anomalous
  Magnetic Moment in an Electromagnetic Wave Field and a Homogeneous
  Magnetic Field
\footnote{Updated version of DESY Report 98--015, February 1998}}

\author{R.A.Melikian \thanks{melikian@jerewan1.YerPhI.am} \\
        Yerevan Physics Institute, \\
        Alikhanian Br. St.2, AM-375036 Yerevan, Armenia \\
        and \\
        D.P.Barber \thanks{mpybar@mail.desy.de} \\
        Deutsches Elektronen-Synchrotron,  DESY ,\ Hamburg, \\
        Notkestrasse 85, 22603 Hamburg, Germany   }

\begin{document}
\maketitle
\begin{abstract}
 The exact solution of the Dirac equation and the spectrum of electron
 quasi-energies in a superposition of the field of a 
 circularly polarized electromagnetic
 wave and a homogeneous magnetic field parallel to the direction of wave
 propagation, are found taking into account the anomalous magnetic
 moment. It is found 
 that taking account of the anomalous magnetic moment removes the spin 
 degeneracy and that for intense fields the levels change radically.
 The shift of the radiation frequency due to the intensity of the wave field
 is found. This shift can be considerable. 
\end{abstract}
\newpage
\tableofcontents
 
\section{Introduction}
 The exact solutions of the Dirac equation in a superposition of 
 a homogeneous magnetic field
 $\vec {\cal B}$ and  a classical monochromatic electromagnetic wave field
 propagating along~$\vec {\cal B}$~ 
 (the so-called Redmond configuration) were
 found for the first time in  \cite{redm-65}.
 In   \cite{berg-83} a transformation was introduced, in analogy with
 the Volkov solution \cite{berest-80}, which reduces  
 the solution of the
 Dirac equation for fields of the Redmond configuration to a product of 
 the solution of a
 Klein-Gordon equation for these fields and a bispinor of  a free particle.
 Wave functions of this form facilitate the physical interpretation,
 calculation and analysis of spin phenomena.
 
 In this paper an exact solution of the Dirac equation and the
 spectrum of particle quasi-energies for  fields of the Redmond
 configuration  is found, taking into
 account the electron anomalous magnetic moment.
 We consider the case of circular wave polarization using the method
 suggested in \cite{berg-83}.
 In addition,
 electron and positron states are distinguished by introducing the
 quantum number~$q_{v}$ which is positive and negative for electrons and
 positrons respectively. 
 
 Furthermore,
 it is shown that by taking into account the electron anomalous magnetic
 moment the degeneracy in the quantum spin states is removed.
  Moreover, owing to the presence of the 
 anomalous magnetic moment and at certain values of other parameters, the
 relative positions of the quasi--energy levels are radically changed. This 
 latter is important for the consideration of spin effects.
 The establishment of the quasi--energy spectrum allows us to obtain
 the photon absorption and radiation spectrum. 
 It is found that the radiation frequency is shifted because of the
 influence of the
 intensity of the wave field and that this shift  can be considerable.

 These theoretical considerations are of direct 
 relevance for various applications: 
 laser acceleration \cite{Milan};
 electron--positron pair creation at high wave field intensities \cite{ole-71};
    fast measurement of the absolute energy of charged 
        unpolarized particles with a relative precision of 
         $10^{-4}$ over a wide energy range up to TeV energies, and so on.
    It is important that during the measurement of the energy of 
 longitudinally spin--polarized electrons by this method, the degree of 
 polarization does not decrease.

\vskip 1cm
 
\section{The solution of the Dirac equation }
 In this section we derive
 the solution of the Dirac equation  in the field of an
 electromagnetic wave and a homogeneous magnetic field taking account of
 the electron anomalous magnetic moment.
 We  describe the external fields in terms of
                                          a classical vector-potential
\begin{eqnarray}
    A(x)= A^{L}(x) + A^{B}(\vec x_{\perp}),
\end{eqnarray}
 where
\begin{eqnarray}
 A^{B}(\vec x_{\perp})=
\Bigl(0;~-~\frac{{\cal B}}{2}{x^2};~\frac{{\cal B}}{2}{x^1};~0\Bigr)
\end{eqnarray}
 is the
 vector-potential of a constant and homogeneous magnetic field 
 $\vec {\cal B}$,
 directed along the $z$ ($\equiv x^3$) axis,
\footnote{In this paper we use two notations in parallel according to 
 convenience: $x^1 \equiv x$, $x^2 \equiv y$, $x^3 \equiv z$.}
  and
\begin{eqnarray}
  A^{L}(u) = \Bigl(0;~-~\frac{{\cal E}}{\omega} \sin\sqrt{2} \omega u;
 {}~\lambda\frac{{\cal E}}{\omega}\cos\sqrt{2}\omega u;~0\Bigr)
\end{eqnarray}
 is the
 vector potential of the monochromatic circularly  polarized wave propagating
 along $\vec {\cal B}$  where
 $u=(x^{0}-x^{3})/\sqrt{2}$ and
                             $\lambda  = \pm 1$  corresponds to the right or
 left polarized wave.
  Lastly,
                      $\omega$ and
                                ${\cal E}$ are the
                                       frequency and amplitude of the
 wave.
 Note that we choose units for which $\hbar=c=1 $.

 We next consider
                 the Dirac equation in the  field
                                       $A(x)$  taking into account the
 electron anomalous magnetic moment. This has the form:
\begin{eqnarray}
\Bigl(\gamma^{\mu} \hat \Pi_{\mu} - m -
 ia|\mu_{0}|F_{\mu \nu} \sigma^{\mu \nu} \Bigr)\Psi_{D}(x) = 0,
\end{eqnarray}
 which can be conveniently rewritten as:
\begin{eqnarray}
\Bigl\{\gamma_{u} \hat P_{v} + \gamma_{v}\hat P_{u} -
 (\vec\gamma {\hat{\vec\Pi}}_{\perp})- m - a|\mu_{0}|[{\cal B}\Sigma_{z}+
 (\vec {\cal H}\vec\Sigma)\gamma_{u}\gamma_{v}]\Bigr\}\Psi_{D}(x) = 0.
\end{eqnarray}
 Here~$F_{\mu \nu}$  is the
                                        external electromagnetic field
 tensor,~ $a = (g-2)/2$~  is the gyromagnetic anomaly and
 $\mu_{0} = e / 2m$~ is  the normal electron magnetic moment.
 
 Furthermore we have defined:
\begin{eqnarray}
\sigma^{\mu \nu}=\frac{1}{2} (\gamma^{\mu} \gamma^{\nu}-
\gamma^{\nu} \gamma^{\mu})=
 (\vec\alpha;i\vec\Sigma);~~~\vec\alpha=\gamma^{0} \vec\gamma;~~~
\gamma_{u}=\frac{\gamma^{0} + \gamma^{3}}{\sqrt{2}};~~~
\gamma_{v}=\frac{\gamma^{0} - \gamma^{3}}{\sqrt{2}};\nonumber
\end{eqnarray}
\begin{eqnarray}
 P_{u}=\frac{P^{0}+P^{3}}{\sqrt{2}};
 {}~~~P_{v}=\frac{P^{0}-P^{3}}{\sqrt{2}};~~
 v=\frac{x^{0}+x^{3}}{\sqrt{2}},\nonumber
\end{eqnarray}
and we use the symbol `~${\hat{~}}$~' to denote differential operators so that
\begin{eqnarray}
\hat P_{v}=i\frac{\partial}{\partial v}=i\partial_{v}=
\frac{i}{\sqrt{2}}\Bigl(\frac{\partial}{\partial x^{0}}+
\frac{\partial}{\partial x^{3}}\Bigr);~~~
\hat P_{u}=i\frac{\partial}{\partial u}=i\partial_{u}=
\frac{i}{\sqrt{2}}\Bigl(\frac{\partial}{\partial x^{0}}-
\frac{\partial}{\partial x^{3}}\Bigr);\nonumber
\end{eqnarray}
\begin{eqnarray}
\hat \Pi_{\mu}=\hat P_{\mu}+|e|A_{\mu}(x);~~{\hat{\vec\Pi}}_{\perp}=
 {\hat{\vec P}}_{\perp}+|e|\vec A_{\perp};~~\gamma^{\mu}\hat P_{\mu}=
\gamma_{u}\hat P_{v}+\gamma_{v}\hat P_{u}-
 (\vec \gamma {\hat{\vec P}}_{\perp}).\nonumber
\end{eqnarray}
 Note that
 $\gamma_{u} \gamma_{v} + \gamma_{v} \gamma_{u}=2$ and
 $\gamma^{2}_{v}=\gamma^{2}_{u}=0$.
 
 In equation (5)  we have used the relation
         $\vec {\cal H}= \vec n \wedge \vec {\cal E} $ for a plane wave, where
      $\vec n= \vec k / |\vec k|$ and where in turn
                                        $\vec k$ is the
                                                       wave vector of the
 electromagnetic wave.
           Then   $(\vec {\cal H}\vec \Sigma)\alpha_{3} = i(\vec {\cal E}\vec
\alpha)$.
 
 Since
 the expression in the curly brackets in (5) does not depend on the
 variable~$v$~ explicitly,          we can separate off the
                                                   $v$-dependence of the
 function~ $\Psi_{D}(x)$. To do this we introduce the replacement:
\begin{eqnarray}
\Psi_{D}(x)=e^{-i(v f_{v}+u f_{u})}\Psi_{f}(u,\vec x_{\perp}),
\end{eqnarray}
 where $f_{u}=(f^{0}+f^{3})/\sqrt{2}$ ~and
                                      ~ $f_{v}=(f^{0}-f^{3})/\sqrt{2}$~
 are arbitrary
            constant numbers whose meaning will emerge later.
   Then $\Psi_{f}(u,\vec x_{\perp})$
 satisfies the equation
\begin{eqnarray}
\Bigl\{\gamma_{u}f_{v}+\gamma_{v}f_{u}+\gamma_{v}i\partial_{u}-
 (\vec\gamma{\hat{\vec \Pi}}_{\perp})- m - a|\mu_{0}|[{\cal B}\Sigma_{z}+
 (\vec {\cal H}\vec \Sigma)\gamma_{u}\gamma_{v}]\Bigl\}
\Psi_{f}(u,\vec x_{\perp})=0.
\end{eqnarray}
 Furthermore, following \cite{berg-83} we make the
                                                 transformation:
\begin{eqnarray}
\Psi_{f}(u,\vec x_{\perp})=\Bigl[1+
\frac{(k\gamma)}{2(kf)}(\vec \gamma{\hat{\vec \Pi}}_{\perp})\Bigr]
\Phi(u,\vec x_{\perp})
\end{eqnarray}
 and since, as shown in \cite{berg-83},
 the function $\Phi$ can be chosen to be an eigenfunction of the spin
  operator
 $\Sigma_{3}$ (projection of $\vec \Sigma$ on to $\vec {\cal B}$ direction) 
 with eigenvalues $\zeta=\pm1$:
\begin{eqnarray}
\Sigma_{3}\Phi=\zeta\Phi,
\end{eqnarray}
 we obtain
\begin{eqnarray}
\Bigl[\gamma_{u}f_{v}+\gamma_{v}f_{u}+\gamma_{v}\Bigl(i\partial_{u}-
\frac{{\hat{\vec \Pi}}_{\perp}^{2}}{2f_{v}}-
\frac{|e|{\cal B}\zeta}{2f_{v}}\Bigr)+
\gamma_{v}(\vec \gamma{\hat{\vec \Pi}}_{\perp})
\frac{a|\mu_{0}|{\cal B}\zeta}{f_{v}}-
\nonumber\\
 -m_{\ast}-
 a|\mu_{0}|\gamma_{v}\gamma_{u}(\vec {\cal H}\vec \Sigma)\Bigr]
\Phi(u,\vec x_{\perp})=0,
\end{eqnarray}
 where  $m_{\ast}$ = $m+a|\mu_{0}|\zeta {\cal B}$.
 
 {}From (10), after some simple manipulations, we obtain
\begin{eqnarray}
\Bigl[i\partial_{u}-\frac{{\hat{\vec \Pi}}_{\perp}^{2}}{2f_{v}}-
\frac{|e|{\cal B}\zeta}{2f_{v}}+
\frac{f^{2}-m_{\ast}^{2}}{2f_{v}}\Bigr]\Phi(u,\vec x_{\perp})=0 ,
\end{eqnarray}
\begin{eqnarray}
\Bigl[(\gamma q)-m\Bigr]\Phi(u,\vec x_{\perp})=0,
\end{eqnarray}
 where
     {}~$f^{2} =2f_{u}f_{v}$ and where we have introduced a new
 four-momentum vector ~$q^{\mu}$~ defined by
\begin{eqnarray}
 q^{\mu}=f^{\mu}\frac{m}{m_{\ast}}-
\frac{k^{\mu}}{2(kf)}\frac{m}{m_{\ast}}(f^{2}-m_{\ast}^{2})
\end{eqnarray}
 with nonvanishing components~$q_{v}=(q^{0}-q^{3})/\sqrt{2}$ ~and
 {}~$q_{u}=(q^{0}+q^{3})/\sqrt{2}$.
 
 It is seen from (11) and (12) that the~$\Phi$~function is a product of
 spinor and spatial parts
\begin{eqnarray}
\Phi(u,\vec x_{\perp})=\chi(u,\vec x_{\perp})U_{q,\zeta} ,
\end{eqnarray}
 where~$U_{q,\zeta}$~is a constant bispinor independent of the variables
 ~$(u,\vec x_{\perp})$ ~satisfying the equations
\begin{eqnarray}
\Bigl[(\gamma q)-m\Bigr]U_{q,\zeta}=0,
\end{eqnarray}
\begin{eqnarray}
\Sigma_{3}U_{q,\zeta}=\zeta U_{q,\zeta}.
\end{eqnarray}
 Because ~$k^{2}=0$~we have from (13)
\begin{eqnarray}
 (kq)=(kf)\frac{m}{m_{*}},~~q_{v}=f_{v}\frac{m}{m_{*}},~~q^{2}=m^{2}.
\end{eqnarray}
 So~$q^{\mu}$~is a free particle 4--momentum with non--zero
 components ~$q_{u}$,~ $q_{v}$ .
 
 In (14)~~$\chi(u,\vec x_{\perp})$~ is a
                                    spin independent function satisfying
 the equation
\begin{eqnarray}
\Bigl(i\partial_{u}-
\frac{{\hat {\vec \Pi}}_{\perp}^{2}}{2f_{v}}-
\frac{|e|{\cal B}\zeta}{2f_{v}}+
\frac{f^{2}-m_{\ast}^{2}}{2f_{v}}\Bigr)\chi(u,\vec x_{\perp})=0.
\end{eqnarray}
 This  can be written in a more convenient form as:
\begin{eqnarray}
\Bigl(i\partial_{u}-
\frac{{\hat {\vec \Pi}}_{\perp}^{2}}{2f_{v}}\Bigr)\chi_{0}(u,\vec x_{\perp})=0
\end{eqnarray}
 by introducing the replacement
\begin{eqnarray}
\chi(u,\vec x_{\perp})=
 \exp\Bigl[-iu\Bigl(\frac{|e|{\cal B}\zeta}{2f_{v}}-
\frac{f^{2}-m_{\ast}^{2}}{2f_{v}}\Bigr)\Bigr]\chi_{0}(u,\vec x_{\perp}).
\end{eqnarray}
 If we assume that~$A(x)\rightarrow 0$~ for ~$x\rightarrow \infty$~, then~
 $\vec {\cal B},\vec {\cal E},\vec {\cal H}\rightarrow 0$,
  {}~$m_{\ast}\rightarrow m$,
 {}~$f^{\mu}\rightarrow q^{\mu}$ and
 {}~$[(\gamma f)-m]U_{f,\zeta}=0$. Then for ~$x\rightarrow \infty$~ 
 {}~$\Psi_{D}(x)$ ~is
 a              solution of the Dirac equation for a free particle,
 and ~$f^{0}$,~$f^{3}$~
 are then the         energy and~$z$~ component of the
                                              electron momentum in the
 absence of external fields.
 
 In order to separate ~$\Psi_{D}(x)$~into electron and positron 
 states we note that
\begin{eqnarray}
 q^{2}=2q_{u}q_{v}=(q^{0})^{2}-(q^{3})^{2}=m^{2} 
\end{eqnarray}
 so that
\begin{eqnarray}
 q^{0}=\frac{2q_{v}^{2}+m^{2}}{2\sqrt{2}q_{v}}.
\end{eqnarray} 
 It is clear
                from (22) that the signs of ~$q^{0}$~
 and ~$q_{v}$~ always coincide, so that
 {}~$q_{v}>0$~(or ~$f_{v}>0$~ because of (17))~- corresponds to an
                                                              electron
 state and~$q_{v}<0$~(or~$f_{v}<0$) ~corresponds to a positron state
\cite{gold-64}.
 
 Taking into account that the potential (1) satisfies the condition
 {}~div$\vec A=0$,~we obtain from (19)
\begin{eqnarray}
 i\partial_{u}\chi_{0}(u,\vec x_{\perp})=
\frac{1}{2f_{v}}\Bigl[{\hat {\vec P}}_{\perp}^{2}+
 2|e|(\vec A {\hat {\vec P}}_{\perp}) +
 e^{2}\vec A_{\perp}^{2}\Bigr]\chi_{0}(u,\vec x_{\perp}).
\end{eqnarray}
 
 The solution of equation (23) is simplified if we transform from the
 coordinate representation into the Fock space (number~  state)
 representation by expressing \\
 {}~$(x;\hat P_{x})$,~$(y;\hat P_{y}$) in terms of  creation
 and annihilation operators
                  ~$(a;a^{+})$,~$(b;b^{+})$~
 for a harmonic oscillator. Then we find from (23)
\begin{eqnarray}
 i\partial_{u}\chi_{0}(u,\vec x_{\perp})=
\Bigl[\frac{\omega_{f}}{2}(a^{+}a + \frac{1}{2})+
\frac{\omega_{f}}{2}(b^{+}b + \frac{1}{2}) +
\frac{\alpha}{\sqrt{2}}(a-ib) + \nonumber\\
 + \frac{\alpha^{*}}{\sqrt{2}}(a^{+}+ib^{+})+
 i\frac{\omega_{f}}{2}(ab^{+} -ba^{+}) +
\frac{e^{2}}{2f_{v}}\Bigl(A^{L}(u)\Bigr)^{2}\Bigr]\chi_{0}(u,\vec x_{\perp}),
\end{eqnarray}
 where
\begin{eqnarray}
\alpha =
\frac{|e|\omega_{f}}{\sqrt{2m\omega_{c}}}\Bigl(A_{y}^{L}(u)-
 iA_{x}^{L}(u)\Bigr)=
\frac{\lambda |e|{\cal E}~\omega_{f}}{\omega\sqrt{2m\omega_{c}}}
 \exp(i\sqrt{2}\lambda \omega u),~
\omega_{f}=\frac{|e|{\cal B}}{f_{v}},~\omega_{c}=\frac{|e|{\cal B}}{m}.
\end{eqnarray}
 The term ~$\frac{e^{2}}{2f_{v}}\Bigl(A^{L}(u)\Bigr)^{2}$~ in square
 brackets in (24) can be eliminated by the replacement:
\begin{eqnarray}
\chi_{0}(u,\vec x_{\perp})=
\chi_{1}(u,\vec x_{\perp})\exp\Bigl[-i\frac{e^{2}}{2f_{v}}
\int\limits_{0}^{u}\Bigl(A^{L}(\tau)\Bigr)^{2}d\tau\Bigr].
\end{eqnarray}
 By substituting (26) into (24) we then obtain an equation for
 {}~$\chi_{1}(u,\vec x_{\perp})$~:
\begin{eqnarray}
 i\partial_{u}\chi_{1}(u,\vec x_{\perp})=
\Bigl[\frac{\omega_{f}}{2}(a^{+}a + \frac{1}{2})+
\frac{\omega_{f}}{2}(b^{+}b +\frac{1}{2})+
\frac{\alpha}{\sqrt{2}}(a-ib)+\nonumber\\
 +\frac{\alpha^{*}}{\sqrt{2}}(a^{+}+ib^{+})+
 i\frac{\omega_{f}}{2}(ab^{+}-ba^{+})\Bigr]\chi_{1}(u,\vec x_{\perp}).
\end{eqnarray}
 We can eliminate the terms linear in
  {}~$a,a^{+},b,b^{+}$~ in the square brackets in (27)
 by using the translation operators ~$D_{\sigma_{a}}=
 \exp\Bigl(\sigma_{a}a^{+}-\sigma_{a}^{*}a\Bigr)$~ and
 {}~$D_{\sigma_{b}}=\exp\Bigl(\sigma_{b}b^{+}-\sigma_{b}^{*}b\Bigr)$~ which
  have
 the following properties:
\begin{eqnarray}
 D_{\sigma_{a}}^{-1}aD_{\sigma_{a}}=a + \sigma_{a},
 {}~~D_{\sigma_{a}}^{-1}a^{+}D_{\sigma_{a}}=a^{+} + \sigma_{a}^{*},
 {}~~D_{\sigma_{a}}^{+}=D_{\sigma_{a}}^{-1},
 {}~~D_{\sigma_{a}}^{-1}D_{\sigma_{a}}=1.\nonumber
\end{eqnarray}
 The operator ~$D_{\sigma_{b}}$ has analogous properties.
 By applying the
   substitution
\begin{eqnarray}
\chi_{1}(u,\vec x_{\perp})=
 D_{\sigma_{a}}D_{\sigma_{b}}\chi_{2}(u,\vec x_{\perp})
\end{eqnarray}
 to (27)
 and using the relation
\begin{eqnarray}
\partial_{u}D_{\sigma_{a}}=
 D_{\sigma_{a}}\Bigl\{(\partial_{u}\sigma_{a})a^{+}-
 (\partial_{u}\sigma_{a}^{\ast})a-
\frac{1}{2}[\sigma_{a}(\partial_{u}\sigma_{a}^{\ast})-
\sigma_{a}^{\ast}(\partial_{u}\sigma_{a})]\Bigr\}\nonumber
\end{eqnarray}
 and an analogous relation for~$D_{\sigma_{b}}$
 we obtain from (27) the equation for $\chi_2$.
 By grouping the terms according to their $a^{+}$,~$a$,~$b^{+}$,~$b$
  content and by
 setting the coefficients of the terms 
 linear in $a^{+}$,~$a$,~$b^{+}$,~$b$ 
 to zero we then obtain the equations:
\begin{eqnarray}
 i \partial_{u} \sigma_{a} = \frac{\omega_{f}}{2} \sigma_{a} +
  \frac{\alpha^{*}}{\sqrt{2}} - i \frac{\omega_{f}}{2} \sigma_{b}
\nonumber
\end{eqnarray}
\begin{eqnarray}
 i \partial_{u} \sigma_{b} = \frac{\omega_{f}}{2} \sigma_{b} +
 i \frac{\alpha^{*}}{\sqrt{2}} + i \frac{\omega_{f}}{2} \sigma_{a}
\nonumber
\end{eqnarray}
 {}From these
 we find that the parameters ~$\sigma_{a}$,~$\sigma_{b}$~ obey the
 condition ~$i\sigma_{a}-\sigma_{b}=const$.~If we now  choose
 {}~$i\sigma_{a}=\sigma_{b}$ then
 {}~for ~$\chi_{2}$ ~and ~$\sigma_{a}$ 
 we obtain the equations:
\begin{eqnarray}
 i\partial_{u}\chi_{2}(u,\vec x_{\perp})=
\Bigl[\frac{\omega_{f}}{2}(a^{+}a + \frac{1}{2})+
\frac{\omega_{f}}{2}(b^{+}b + \frac{1}{2})+\nonumber\\
 +\frac{1}{\sqrt{2}}(\alpha\sigma_{a}+\alpha^{*}\sigma_{a}^{*})+
 i\frac{\omega_{f}}{2}(ab^{+} -
 ba^{+})\Bigr]\chi_{2}(u,\vec x_{\perp}),
\end{eqnarray}
\begin{eqnarray}
\partial_{u}  {\sigma_{a}}+i\omega_{f}\sigma_{a}=
 - i\frac{\alpha^{*}}{\sqrt{2}}.
\end{eqnarray} ~We can eliminate the
 term~$\frac{1}{\sqrt{2}}(\alpha\sigma_{a} + \alpha^{\ast}\sigma_{a}^{\ast})$~
 in (29) by the replacement
\begin{eqnarray}
\chi_{2}(u,\vec x_{\perp})=
 \exp\Bigl[-i\sqrt{2}\int\limits_{0}^{u}Re\Bigl(\alpha(\tau)\sigma(\tau)
\Bigr)d\tau\Bigr]\chi_{3}(u,\vec x_{\perp})
\end{eqnarray}
 and by substituting (31) in (29) we obtain an equation
 for~$\chi_{3}(u,\vec x_{\perp})$:
\begin{eqnarray}
 i\partial_{u}\chi_{3}(u,\vec x_{\perp})=
\Bigl[\frac{\omega_{f}}{2}(a^{+}a +\frac{1}{2})+
\frac{\omega_{f}}{2}(b^{+}b + \frac{1}{2})+
 i\frac{\omega_{f}}{2}(ab^{+}-ba^{+})\Bigr]\chi_{3}(u,\vec x_{\perp}).
\end{eqnarray}
 If now in (32) we return to  a coordinate representation by introducing
 cylindrical coordinates~($\varrho,\varphi$), we find
\begin{eqnarray}
 i\partial_{u}\chi_{3}(u,\vec x_{\perp})=
\Bigl[\frac{\omega_{f}^{2}f_{v}}{8}\varrho^{2}-
\frac{1}{2f_{v}}\Bigl(\frac{\partial^{2}}{\partial\varrho^{2}}+
\frac{1}{\varrho}\frac{\partial}{\partial\varrho}+
\frac{1}{\varrho^{2}}\frac{\partial^{2}}{\partial\varphi^{2}}\Bigr)-
 i\frac{\omega_{f}}{2}\frac{\partial}{\partial\varphi}\Bigr]
\chi_{3}(u,\vec x_{\perp}).
\end{eqnarray}
 Since the expression in the square bracket in (33) does not
                                                        depend on the
 variable ~$u$ ~we can factor ~$\chi_{3}(u,\vec x_{\perp})$
 into the form
\begin{eqnarray}
\chi_{3}(u,\vec x_{\perp})=
 \exp(-iE_{n}u)\phi_{n,s}(\varrho,\varphi),
\end{eqnarray}
 where ~$\phi_{n,s}(\varrho,\varphi)$ ~obeys the equation
\begin{eqnarray}
\Bigl[\frac{\omega_{f}^{2}f_{v}}{8}\varrho^{2} -
\frac{1}{2f_{v}}\Big(\frac{\partial^{2}}{\partial\varrho^{2}} +
\frac{1}{\varrho}\frac{\partial}{\partial\varrho} +
\frac{1}{\varrho^{2}}\frac{\partial^{2}}{\partial\varphi^{2}}\Bigr)-
 i\frac{\omega_{f}}{2}\frac{\partial}{\partial\varphi}
 - E_{n}\Bigr]\phi_{n,s}(\varrho,\varphi)=0
\end{eqnarray}
 and the parameter ~$E_{n}$ ~will be defined below.
 Furthermore,
          we can factor ~$\phi_{n,s}(\varrho,\varphi)$ ~into terms
 depending on
 {}~$\varrho$ and $\varphi$:
\begin{eqnarray}
\phi_{n,s}(\varrho,\varphi) = \exp(il\varphi)R(\varrho).
\end{eqnarray}
 By introducing the dimensionless variable ~$\eta =
\frac{|e|{\cal B}}{2}\varrho^{2}$
 {}~ the equation for ~$R$ ~can be written in the form
\begin{eqnarray}
\eta R''+ R'+\Bigl(\varepsilon -\frac{\eta}{4}-\frac{l^{2}}{4\eta}\Bigr)R =
 0 ,
\end{eqnarray}
 where ~$\varepsilon = \frac{E_n}{\omega_{f}} - \frac{l}{2}$~ and the prime
 denotes differentiation with respect to ~$\eta$. ~Equation
 (37) has the same form as the equation for the radial wave function of
 the Schrodinger equation in a homogeneous magnetic field ~$\vec {\cal B}$~ for
 states where the electron possesses definite momentum and angular
 momentum values along the direction of the field
 {}~$\vec {\cal B}$~\cite{land-89}. The solution of the equation (37) can be
 expressed in terms of Laguerre functions \cite{land-89}, \cite{sok-83}:
\begin{eqnarray}
 R(\eta)=N_{n,s}I_{n,s}(\eta),
\end{eqnarray}
 where
\begin{eqnarray}
 I_{n,s}(\eta)=
\frac{1}{\sqrt{n!s!}}e^{-\frac{\eta}{2}}\eta^{\frac{|l|}{2}}
 L_{s}^{|l|}(\eta) ,\nonumber
\end{eqnarray}
 $N_{n,s}$~is a normalization factor defined below and
 {}~$L_{s}^{|l|}$~ are associated
             Laguerre polynomials. The constraint that the
 function ~$R$~ must vanish as
                   ~$\eta\rightarrow\infty$~ means that
                         $\varepsilon  -\frac{|l|+1}{2} = s$,~
 where~$s=0,1,2,...$~is a radial quantum number. This in turn allows us
 to obtain the values:
\begin{eqnarray}
 E_{n}=\omega_{f}\Bigl(s +\frac{l+|l|}{2} +\frac{1}{2}\Bigr)=
\frac{|e|{\cal B}}{f_{v}}\Bigl(n +\frac{1}{2}\Bigr)
\end{eqnarray}
 for ~$E_n$.~Here~$n = s +\frac{l}{2} +\frac{|l|}{2}= 0,1,2,...$~  is the
 principle
 quantum number and ~$l$~ is the value of the
                      component of angular momentum
 along~$\vec {\cal B}$ (for ~$l>0$~we have~$l = n - s$;~for  ~$l<0$ ~we have
 {}~$n - s = 0$).
 
 Substituting  (26) into (20) and using the
 relations (28), (31), (34), (36), (38), (39) we now obtain the solution
 of equation  (18), namely:
\begin{eqnarray}
\chi(u,\vec x_{\perp})=
 \exp\Bigl\{-i\frac{u}{2f_{v}}\Bigl[|e|{\cal B}(2n+1+\zeta)+m_{\ast}^{2}-
 m^{2}\Bigr]-\nonumber\\
 - i\frac{e^{2}}{2f_{v}}\int\limits_{0}^{u}\Bigl(A^{L}(\tau)\Bigr)^{2}d\tau-
 i\sqrt{2}\int\limits_{0}^{u}Re\Bigl(\alpha(\tau)\sigma_{a}(\tau)\Bigr)d\tau
\Bigr\}F_{n,s}(u,\vec x_{\perp}) ,
\end{eqnarray}
 where
\begin{eqnarray}
 F_{n,s}(u,\vec x_{\perp})=D_{\sigma_{a}}D_{\sigma_{b}}\phi_{n,s}(x,y),
 {}~~~\phi_{n,s}(x,y)=N_{n,s}e^{il\varphi}I_{n,s}(\eta).
\end{eqnarray}
 We are mainly interested in the nonresonant case 
 $((kf)-\lambda |e|{\cal B} \not= 0)$) and then 
 the parameter ~$\sigma_{a}$, ~found from equation (30)  has the
  form
\begin{eqnarray}
\sigma_{a}=
\frac{|e|{\cal B}}{2\sqrt{m\omega_{c}}}\frac{|e|{\cal E}}{\omega}
\frac{\exp(-i\sqrt{2}\lambda \omega u)}{(kf)-\lambda |e|{\cal B}}.
\end{eqnarray}
 Exactly at the resonance~$(kf) - |e|{\cal B} = 0$ ~we
 have~$Re(\alpha\sigma_{a})= 0$.
 {}~Expressing the operators ~$D_{\sigma_{a}}$ and
                                             ~$D_{\sigma_{b}}$ ~in terms
 of
 {}~$(x;\hat P_{x})$, ~$(y;\hat P_{y})$ ~and using the
 expressions (25), (42) we obtain
\begin{eqnarray}
 F_{n,s}(u,\vec x_{\perp})=
 e^{i\alpha_{1}(-x(\sin\sqrt{2}\lambda \omega u)+
 y(\cos\sqrt{2}\lambda \omega u))}
e^{-i\beta_{1}((\cos\sqrt{2}\lambda \omega u)\hat
 P_{x}+ (\sin\sqrt{2}\lambda \omega u)\hat P_{y})}\phi_{n,s}(x,y)= \nonumber\\
 =\exp\Bigl[i\alpha_{1}(-x\sin\sqrt{2}\lambda \omega u+
 y\cos\sqrt{2}\lambda \omega u)\Bigr]\phi_{n,s}(x-\beta_{1}\cos\sqrt{2}\lambda
  \omega u; ~y-
\beta_{1}\sin\sqrt{2}\lambda \omega u)  \nonumber \\
\end{eqnarray}
 where
\begin{eqnarray}
\beta_{1}=\frac{|e|{\cal E}}{\omega}\frac{1}{(kf)-\lambda |e|{\cal B}}~,
 {}~~~\alpha_{1}=
\frac{|e|{\cal B}}{2}\beta_{1}.
\end{eqnarray}
 To calculate the operators ~$D_{\sigma_{a}}$, ~$D_{\sigma_{b}}$we used
 {}~the Weyl identity
\begin{eqnarray}
 e^{\hat A+\hat B}=e^{-\frac{1}{2}[\hat A,\hat B]}e^{\hat A}e^{\hat B} ,
\nonumber
\end{eqnarray}
 which is valid for operators ~$\hat A$, $\hat B$ ~satisfying the equality
 {}~$[\hat A,[\hat A,\hat B]] = [\hat B,[\hat A,\hat B]]=0$.
 
 So from (6),(8),(14),(40) and using (3),(25),(42) we finally obtain the
 following
 expression for the electron wave function
\begin{eqnarray}
\Psi_{D}(x)=e^{-i(Qx)}\Bigl[1+
\frac{(k\gamma)}{2(kf)}(\vec\gamma{\hat{\vec\Pi}}_{\perp})\Bigr]
 U_{q,\zeta}F_{n,s}(u,\vec x_{\perp}).
\end{eqnarray}
 where
\begin{eqnarray}
 Q^{\mu}=q^{\mu}\frac{m_{\ast}}{m}+
\frac{k^{\mu}}{2(kQ)}\Bigl[|e|{\cal B}(2n+1+\zeta)+
\frac{\xi^{2}m^{2}(kQ)}{(kQ)-\lambda |e|{\cal B}}\Bigr] ,
\end{eqnarray}
 and where~$\xi=|e|{\cal E} / m \omega$~is the intensity parameter of the
 electromagnetic wave field.
 The wave function (45) satisfies the periodicity conditions
\begin{eqnarray}
 {{\Psi}_D}(t+T) = e^{-i{Q^0}T} {{\Psi}_D}(t)~,~~~
 {{\Psi}_D}(z+\lambda_{w}) = 
   e^{-i{Q^3}\lambda_{w}} {{\Psi}_D}(z) ,  \nonumber 
  \end{eqnarray}
 where $T=\frac{2\pi}{\omega}$ is the period of the oscillations 
   and ~$\lambda_{w}$~ is the  wavelength. 
   Therefore  according to the definition of
 quasi-momentum  \cite{zeld-66},\cite{nik-64},~$Q^{\mu}$ is a particle 
 quasi--momentum
  {}~with
 components ~$Q^{0}$ ~and ~$Q^{3}$.
 
 The normalization factor~$N_{n,s}$ ~is obtained from the condition
\begin{eqnarray}
\int\Psi_{D}^{\ast}(x)\Psi_{D}(x)d^{3}x=1.
\end{eqnarray}
 Noticing that
\begin{eqnarray}
\Psi^{\ast}_D=
\bar\Psi\gamma^{0},~~~\bar\Psi_D=e^{i(Qx)}\bar U_{q,\zeta}\Bigl[1-
\frac{\gamma_{v}}{2f_{v}}(\vec\gamma{\hat{\vec\Pi}}_{\perp}^{\ast})\Bigr]
 F_{n,s}^{\ast}(u,\vec x_{\perp})
\end{eqnarray}
 and substituting~(45),~(48)~into~(47)~we obtain
\begin{eqnarray}
 (\bar U_{q}\gamma^{0}U_{q})\int F_{n,s}^{\ast}F_{n,s}d^{3}x-
\frac{1}{2f_{v}}(\bar U_{q}\gamma_{v}\gamma^{\alpha}\gamma^{0}U_{q})
\int(\hat \Pi^{\ast\alpha}F_{n,s}^{\ast})F_{n,s}d^{3}x+\nonumber\\
 + \frac{1}{2f_{v}}(\bar U_{q}\gamma^{0}\gamma_{v}\gamma^{\alpha}U_{q})
\int F_{n,s}^{\ast}(\hat\Pi^{\alpha}F_{n,s})d^{3}x-\nonumber\\
 - \frac{1}{(2f_{v})^{2}}(\bar U_{q}\gamma_{v}\gamma^{\alpha}\gamma^{0}
\gamma_{v}\gamma^{\beta}U_{q})\int(\hat\Pi^{\ast\alpha}F_{n,s}^{\ast})
 (\hat\Pi^{\beta}F_{n,s})d^{3}x=1.\\
\alpha,\beta=1,2.\nonumber
\end{eqnarray}
 Assuming the normalization ~$\bar U_{q}U_{q}= 2m$ ~from (15) we find
\begin{eqnarray}
\bar U_{q}\gamma^{\mu}U_{q}=2q^{\mu} ,~~~~~~~~~~~~~\mu=0,3.
\end{eqnarray}
 By using the expressions~(41),~(43),~(50) and taking the
 formula \cite{sok-83}
\begin{eqnarray}
\int\limits_{0}^{\infty}I_{n,s}^{2}(\eta)d\eta = 1\nonumber
\end{eqnarray}
 into account we find that the first term  of the left part of the expression
 (49) is given by
\begin{eqnarray}
 2q^{0}\int F_{n,s}^{\ast}F_{n,s}d^{3}x=
 2q^{0}\frac{N_{n,s}^{2}} {|e|{\cal B}}\int\limits_{-L_{z}/2}^{+L_{z}/2}dz
\int\limits_{0}^{2\pi}d\varphi\int\limits_{0}^{\infty}I_{n,s}^{2}(\eta)d\eta=
 4\pi q^{0}L_{z}\frac{N_{n,s}^{2}}{|e|{\cal B}}.
\end{eqnarray}
 If we transform the second integral in~(49) using the relation
\begin{eqnarray}
\int(\hat\Pi^{\ast\alpha}F_{n,s}^{\ast})F_{n,s}d^{3}x=
\int F_{n,s}^{\ast}(\hat\Pi^{\alpha}F_{n,s})d^{3}x ,\nonumber
\end{eqnarray}
 and then the relation
\begin{eqnarray}
 {}~-\bar U_{q}\gamma_{v}\gamma^{\alpha}\gamma^{0}U_{q}+
\bar U_{q}\gamma^{0}\gamma_{v}\gamma^{\alpha}U_{q}=
\frac{1}{\sqrt{2}}
\bar U_{q}\Bigl[\gamma_{v}(\gamma_{u}+
\gamma_{v})\gamma^{\alpha}+
 (\gamma_{u}+\gamma_{v})\gamma_{v}\gamma^{\alpha}\Bigr]U_{q}=\nonumber\\
 =\frac{1}{\sqrt{2}}\bar U_{q}(\gamma_{v}\gamma_{u}+
\gamma_{u}\gamma_{v})\gamma^{\alpha}U_{q}=
\sqrt{2}\bar U_{q}\gamma^{\alpha}U_{q}=0 ~\nonumber
\end{eqnarray}
 we see that the second and the third terms of the left hand 
 side of the equality
 (49)~are cancelled.
 Noticing that
\begin{eqnarray}
\bar U_{q}\gamma_{v}\gamma^{\alpha}\gamma^{0}\gamma_{v}\gamma^{\beta}U_{q}=
 -\frac{1}{\sqrt{2}}\bar U_{q}\gamma^{\alpha}\gamma_{v}(\gamma_{u}+
\gamma_{v})\gamma_{v}\gamma^{\beta}U_{q}=
 -\sqrt{2}\bar U_{q}\gamma^{\alpha}\gamma_{v}\gamma^{\beta}U_{q}=\nonumber\\
=\sqrt{2}\bar U_{q}\gamma_{v}\gamma^{\alpha}\gamma^{\beta}U_{q}\nonumber
\end{eqnarray}
 and transforming the fourth integral in~(49) by using the relation
\begin{eqnarray}
\int(\hat\Pi^{\ast\alpha}F_{n,s}^{\ast})(\hat\Pi^{\beta}F_{n,s})d^{3}x =
\int F_{n,s}^{\ast}(\hat\Pi^{\alpha}\hat\Pi^{\beta}F_{n,s})d^{3}x\nonumber
\end{eqnarray}
 we can write the last term on the left hand side of~(49) in the form
\begin{eqnarray}
 -\frac{\sqrt{2}}{(2f_{v})^{2}}
 (\bar U_{q}\gamma_{v}\gamma^{\alpha}\gamma^{\beta}U_{q})\int F_{n,s}^{\ast}
 (\hat\Pi^{\alpha}\hat\Pi^{\beta}F_{n,s})d^{3}x=~~~~~~~~~~~~\nonumber\\
 =\frac{\sqrt{2}}{(2f_{v})^{2}}(\bar U_{q}\gamma_{v}U_{q})
\int F_{n,s}^{\ast}\Bigl[{\hat{\vec\Pi}}^{2}_{\perp}+
 i\zeta(\hat\Pi^{1}\hat\Pi^{2}-
\hat\Pi^{2}\hat\Pi^{1})\Bigr]F_{n,s}d^{3}x=\nonumber\\
 =\frac{\sqrt{2}~2\pi L_{z}}{(2f_{v})^{2}} 
\frac{N_{n,s}^{2}} {|e|{\cal B}}2q_{v}
 |e|{\cal B}\zeta +\frac{\sqrt{2}}{(2f_{v})^{2}}2q_{v}
\int F_{n,s}^{\ast}({\hat{\vec\Pi}}_{\perp}^{2}F_{n,s})d^{3}x.
\end{eqnarray}
 Here we  used the relations~(51), the relations
 {}~~$\hat\Pi^{1}\hat\Pi^{2}-\hat\Pi^{2}\hat\Pi^{1}= -i|e|{\cal B}$, \\
 $\Sigma_{3}U_{q}=i\gamma^{1}\gamma^{2}U_{q}=\zeta U_{q}$~ and
 {}~$\bar U_{q}\gamma_{v}U_{q}=2q_{v}$ ~(following  from~(15)~and~(50)).
 
 {}From (18), (40) we obtain the relation
\begin{eqnarray}
\Bigl[|e|{\cal B}(2n+1)+\frac{m^{2}\xi^{2}(kf)}{(kf)-\lambda
  |e|{\cal B}}+2if_{v}\partial_{u}\Bigr]
 F_{n,s}(u,\vec x_{\perp})={\hat{\vec\Pi}}_{\perp}^{2}F_{n,s}(u,\vec x_{\perp}).
\end{eqnarray}
 Substituting~${\hat{\vec\Pi}}_{\perp}^{2}F_{n,s}$~from~(53)~into~(52)~and
 using~(43),~(51)~we find
\begin{eqnarray}
\int F_{n,s}^{\ast}({\hat{\vec\Pi}}_{\perp}^{2}F_{n,s})d^{3}x=
\Bigl[|e|{\cal B}(2n+1)+\frac{m^{2}\xi^{2}(kf)}{(kf)-\lambda 
 |e|{\cal B}}\Bigr]
\int F_{n,s}^{\ast}F_{n,s}d^{3}x+\nonumber\\
 +i\int F_{n,s}^{\ast}(\partial_{u}F_{n,s})d^{3}x=
 2\pi L_{z}\frac{N_{n,s}^{2}}{|e|{\cal B}}
\Bigl[|e|{\cal B}(2n+1)+\frac{m^{2}\xi^{2}(kf)^{2}}{((kf)-\lambda 
 |e|{\cal B})^{2}}\Bigr].~~~
\end{eqnarray}
 Substituting~(51),~(52),~(54)~into~(49)~we obtain
\begin{eqnarray}
 N_{n,s}=
\Bigl(\frac{|e|{\cal B}}{4\pi L_{z}q^{0}}\Bigr)^{\frac{1}{2}}\Bigl\{1+
\frac{(q^{0}+q^{3})}
 {2q^{0}}\Bigl[\frac{|e|{\cal B}}{m_{\ast}^{2}}(2n+1+\zeta)+
\frac{\xi^{2}(kq)^{2}}{((kq)\frac{m_{\ast}}{m}-\lambda |e|{\cal B})^{2}}\Bigr]
\Bigr\}^{-\frac{1}{2}}.
\end{eqnarray}
\vskip 1cm
 
\section{The electron and positron quasi--energy spectra}
 In the approximation that
                       {}~$m_{\ast}^{2}\simeq m^{2}+|e|{\cal B}\zeta a$ ~(i.e.
 neglecting
 the term $(|e|{\cal B}a / 2m)^{2}$ because it is  small) from (46)
 we obtain a dispersion equation
\begin{eqnarray}
 Q^{2}=(Q^{0})^{2}-(Q^{3})^{2}=m^{2}+|e|{\cal B}(2n+1+\zeta +\zeta a)+
\frac{m^{2}\xi^{2}(kQ)}{(kQ)-\lambda |e|{\cal B}}.
\end{eqnarray}
 In particular, as ~$\vec {\cal B}\rightarrow 0$~ from (46) we obtain  the
 known \cite{berest-80} value of the
 4-quasi-momentum (the average over time of the
 kinetic 4-momentum) of an electron
 in a plane wave electromagnetic field:
\begin{eqnarray}
 Q^{\mu}=q^{\mu}+\frac{m^{2}\xi^{2}}{2(kq)}k^{\mu}.
\end{eqnarray}
 For the case when ~$\xi\rightarrow 0$,~ (56) gives us the usual
 expression for  the energy of an
 electron with anomalous magnetic moment in a   constant magnetic
 field (in 
 the approximation ~${\cal B}\ll {\cal B}_{c}=m^{2}/|e|$)\cite{conn-68}:
\begin{eqnarray}
 (Q^{0})^{2}=(Q^{3})^{2}+m^{2}+|e|{\cal B}(2n+1+\zeta+\zeta a).
\end{eqnarray}
 
 Let us now  find the relative positions  of the quasi--energy   levels
 {}~$Q_{n,\zeta}^{0}$ for a
 circularly polarized wave. Due to the condition $k^{2}= 0$, ~from
 (46) we have
\begin{eqnarray}
 (kQ)=(kq)\Bigl(1+\zeta\frac{a\omega_{c}}{2m}\Bigr)
\end{eqnarray}
 and the following
     expressions for the quasi--energy ~$Q_{n,\zeta}^{0}$ ~and
 quasi--momentum~ $Q_{n,\zeta}^{3}$:
\begin{eqnarray}
 Q_{n,\zeta}^{0}=q^{0}\Bigl(1+\zeta G \Bigr)+
\frac{\gamma \omega_{c}(2n+1+\zeta)}{(1+\zeta G)}
 +\frac{\gamma  m \xi^{2}} 
  {1-\lambda\frac{2 \gamma \omega_{c}}{\omega }+
\zeta G},
\end{eqnarray}
\begin{eqnarray}
 Q_{n,\zeta}^{3}=q^{3}(1 + \zeta G) +
\frac{\gamma \omega_{c}(2n+1+\zeta)}{(1+\zeta G)}
 +\frac{\gamma  m \xi^{2}} 
  {1-\lambda\frac{2 \gamma \omega_{c}}{\omega }+
\zeta G}~~,~
\end{eqnarray}
 where for brevity we use $G=a\omega_{c}/2m$ and 
 $\gamma =(q^{0}+q^{3})/{2 m}$.
 
 Hence, we can see that~ $Q_{n,\zeta}^{3}$ ~depends on the
                                                       quantum numbers
 {}~$n$ ~and~ $\zeta$,~and the
                         difference between the quasi--energy
                                                              levels with
 similar ~$\zeta$ ~is 
\begin{eqnarray}
 Q_{n,\zeta}^{0}-Q_{n',\zeta}^{0}=Q_{n,\zeta}^{3}-Q_{n',\zeta}^{3}=
\frac{2 \gamma \omega_{c}(n-n')}{(1+\zeta G)},
\nonumber
\end{eqnarray}
 i.e. the transition~ $Q_{n,\zeta}^{0}\rightarrow Q_{n',\zeta}^{0}$ ~is
 associated with  the transition between the
 states with different values of the $z$~ component of the quasi--momentum.
 Since $G$ is usually $\ll 1$ (near a resonance we have: 
 $G= \frac{a}{4} \frac{\hbar \omega}{m c^2} \frac{1}{\gamma} 
\approx \frac{10^{-9}}{4}\frac{1}{\gamma} \ll 1 $)
 then in the approximation that we neglect $G$ 
 the 
 difference between the neighbouring quasi-energy levels with
 similar ~$\zeta$ ~is approximately
\begin{eqnarray}
 Q_{n+1,\zeta}^{0}-Q_{n,\zeta}^{0}\simeq 2\gamma\omega_{c}.
\end{eqnarray}
 We will now find the relative positions of quasi--energy  levels
 corresponding to opposite spin directions. If
\begin{eqnarray}
 G \ll |\delta| =
 |1-\lambda \frac{2 \gamma \omega_{c}}{\omega}|  \nonumber
\end{eqnarray}
 {}~from (60),(61) we obtain the approximate relationship:
\begin{eqnarray}
 Q_{0,+1}^{0}-Q_{0,-1}^{0}=Q_{0,+1}^{3}-Q_{0,-1}^{3}=
 2\gamma\omega_{c}(1+\frac{a}{2}-a\frac{\xi^{2}}{\delta^{2}}).
\end{eqnarray}
 The quantity $\delta$ describes the detuning of the frequencies
 $\omega$ and $2 \gamma {\omega}_c$.
 In the  case of weak wave fields i.e. when 
    {}~$\xi^{2}/\delta^{2}\ll 1$, ~(63) gives us
     ~$Q_{0,+1}^{0}-Q_{0,-1}=2\gamma\omega_{c}(1+\frac{a}{2})$. Using
 this relation and (62) we find the quasi--energy spectrum shown schematically
 in Fig 1.
 This shows that inclusion of the electron anomalous
 magnetic moment removes the degeneracy between the levels
  $Q^0_{n,+1}$ and
 $Q^0_{n+1, -1}$.
 
 For intense wave fields, i.e. when ~$a\xi^{2}/\delta^{2}\gg 1$,
 {}from (63) it follows that the
 relative positions of the quasi--energy   levels with opposite spin
 direction
 change radically (Fig.2). On some levels
  in the
 lower part of the spectrum the  electrons can have only  one spin direction
 and the number of these levels is
 {}~$(Q_{0,-1}^{0}-Q_{0,+1}^{0})/2\gamma\omega_{c}$).
 
 {}From these considerations of the quasi--energy  spectrum it is seen
 that inclusion
 of the anomalous magnetic moment in the electron quasi--energy
 spectrum is
 essential at all values of ~$\xi^{2}/\delta^{2}$.
 
 Note  that by summing the values of  ~ $Q^{\mu}$
 found for ~$\lambda = +1$ ~and ~$\lambda =-1$ ~from (46) we find the following
 expression for the particle quasi-energy in  linearly  polarized waves:
\begin{eqnarray}
 Q^{\mu}=q^{\mu}\frac{m_{\ast}}{m}+
\frac{k^{\mu}}{2(kQ)}\Bigl[|e|{\cal B}(2n+1+\zeta)+
\frac{m^{2}\xi^{2}(kQ)^{2}}{(kQ)^{2}-(e{\cal B})^{2}}\Bigr].\nonumber
\end{eqnarray}
 Hence, in the approximation that
                             ~$m_{\ast}^{2}\simeq m^{2}+|e|{\cal B}\zeta a$,~we
 obtain the  dispersion equation
\begin{eqnarray}
 Q^{2}=(Q^{0})^{2}-(Q^{3})^{2}=m^{2}+|e|{\cal B}(2n+1+\zeta +\zeta a)+
\frac{m^{2}\xi^{2}(kQ)^{2}}{(kQ)^{2}-(e{\cal B})^{2}} \nonumber
\end{eqnarray}
 for a   linearly  polarized wave (taking account of the electron anomalous
 magnetic moment) which agrees with the analogous expression 
(except for the absent anomalous magnetic moment) in
\cite{ole-71}.
 
\vskip 1cm
\section{The photon absorption and radiation spectrum}
 For absorption  of a photon with 4--momentum $k^{\mu}(\omega; \vec k)$
 by an electron moving at an angle $\theta =0$ to the $z$--axis, the 
 energy--momentum conservation law reads as:
\begin{eqnarray}
 Q^{\mu} + k^{\mu} = Q'^{\mu}.
\end{eqnarray}
 {}From this and using $Q^{\mu}$ from (46) we find that the photon absorption 
 frequency $\omega$  is: 
\begin{eqnarray}
\omega = 
\frac{\gamma {\omega}_c [2(n'-n) + ({\zeta}'-\zeta)(1+ a)]}
 {1+ {\zeta} G}.
\end{eqnarray}
 For $G = 0$  and without spin flip
 the frequency $\omega$ coincides with that found earlier by
 classical methods.
 
 We see from (65) that if there is no spin flip, the frequencies of
  the  $Q^0$ spectrum are  equally  spaced. Then 
 the electron can resonantly absorb a succession   of 
 photons of the same frequency and accelerate. 
 
 For radiation of a photon with 4--momentum 
 ${k'}({\omega}'; {\vec k}')$
 by an electron the energy--momentum  conservation equation is given by: 
\begin{eqnarray}
 Q^{\mu} + \nu k^{\mu} = {Q'}^{\mu} + {k'}^{\mu}
\end{eqnarray}
 where the coefficient  $\nu$ denotes the number of photons 
  absorbed from the wave field.
 
 {}From (66) and using $Q^{\mu}$ from (46) we find a cubic equation for the 
 frequency spectrum ${\omega}'$ namely:
\begin{eqnarray}
 g_3 {{\omega}'}^3 - g_2 {{\omega}'}^2 + g_1 {\omega}' - g_0 =0 ~,
\end{eqnarray}
 where 
\begin{eqnarray}
 g_0 \equiv \frac{\nu\omega}{\gamma} + {\omega}_c
 [2(n-n') + ({\zeta}-{\zeta}')(1+ a)]~,~ \nonumber
\end{eqnarray}
\begin{eqnarray}
 g_1 \equiv \frac{(1 + \cos \theta)(1+\zeta G)}{2 \gamma} +
     2 \gamma (1 - \cos \theta)
\Bigl\{1 + \zeta G +\frac{\omega_c}{m} \frac{(2n+1+\zeta)}{1 + \zeta G}+
                                                              \nonumber \\  
    \frac{\omega_c}{m} \frac{[2(n-n') + ({\zeta}-{\zeta}')(1+ a)]}
       {\delta + \zeta G}+ 
       \frac{\xi^2 (1 + \zeta G)}{(\delta + \zeta G)^2}
       + \frac{\nu\omega}{m \gamma}(1+\frac{1+\zeta G}
       {\delta +\zeta G}) \Bigr\}~~,\nonumber
\end{eqnarray}
\begin{eqnarray}
 g_2 \equiv \frac{{\sin}^2 \theta}{m} + \frac{(1- \cos\theta)}
            {m ( \delta + \zeta G)} 
            \Bigl\{4 {\gamma}^2 (1- \cos \theta)
            [1+\zeta G +\frac{\omega_c}{m}\frac{(2 n +1 +\zeta)}{1 +
            \zeta G}+\nonumber \\
            \frac{\xi^2}{\delta+\zeta G}] +
            (1+\cos \theta)(1+\zeta G) +\frac{4 \gamma \nu \omega }{m}
            (1-\cos \theta) \Bigr\}~~,\nonumber
\end{eqnarray}
\begin{eqnarray}
 g_3 \equiv \frac{2 \gamma {\sin}^2\theta(1-\cos \theta)}{(\delta+\zeta G)
  m^2} ~~. \nonumber
\end{eqnarray}
 
 For photons radiated along the $z$--axis ($\cos \theta = 1$) we find from 
 (67) that:
\begin{eqnarray}
 {\omega}' = \frac{g_0}{g_1}=\frac{\nu\omega+ \gamma
             \omega_c[2(n-n')+(\zeta-{\zeta}')(1+a)]}{1+\zeta G}~~ .
\end{eqnarray}
 
 For the radiation of photons at optical and lower frequencies the quantum
 corrections in (67) can be neglected and then, for an arbitrary $\theta$,
 we obtain:
\begin{eqnarray}
 {\omega}'=\frac{\nu\omega+ 2 \gamma \omega_c(n-n')}
         {\frac{1+\cos\theta}{2} + 2 {\gamma}^2 (1-\cos \theta)
         (1+\frac{\xi^2}{\delta^2})}~~ .
\end{eqnarray}
 
 We see from (69) that when $\cos \theta \ne 1$ the term $\xi^2/\delta ^2$ 
 would lead to a considerable 
 change, namely a decrease, in the radiation 
 frequency ${\omega}'$  compared with the scattering of an electromagnetic 
 wave on free electrons --- especially near resonance when $\delta \ll 1$.
   
 Because the intensity of the radiation increases near resonance like
 $\xi^2/\delta^2$ \cite{Bagrov},
  the resonance condition $\omega=2 \gamma \omega_c$ 
  can
  be used for measurement of the particle energy and then
 the  shift of the radiation frequency ${\omega}'$ (69) 
 compared to the incident laser frequency $\omega$ allows incident and
 radiated photons to be distinguished \cite{spin98}.
 
 Note that the resonance condition $\omega = 2 \gamma \omega_c$ is easy
 to realise using existing powerful lasers (with $\xi$ up to $1$) for
 reasonable values of the parameters 
 $\gamma$ and $\omega_c$. For example at an electron
 energy of $500~GeV$  and a laser wavelength 
 $\lambda_{w}= 0.248 \mu m$ ($\hbar \omega = 5 eV$) 
  {} from a KrF laser,
 one needs a field ${\cal B} = 220~Gs$.
\vskip 1cm
\section{Summary}
 We have found the wave functions (45) and electron quasi-energy
 spectrum  (46) and (56) in a superposition of a homogeneous magnetic field~ 
 $\vec {\cal B}$ ~and a
 classical circularly polarized wave propagating along~ $\vec {\cal B}$~ for an
 electron with anomalous magnetic moment.
 The factorized wave functions (45) facilitate
 calculations for  processes involving particles with spin since the
 bispinor~$U_{q,\zeta}$ ~satisfies the equation for a free particle.
 
 Taking account of the electron anomalous magnetic moment removes the
 degeneracy of the quantum spin states. For intense wave fields, i.e.~
 when
 $a\xi^{2}/\delta^{2} \gg 1$ ~the relative positions of the
 quasi--energy levels with opposite spin direction change radically (Fig. 2).
 
 We have considered the photon absorption and radiation spectrum and have found 
 a dependence of the shift of the radiation frequency  on the 
 intensity of the wave field. 
 
 In addition we suggest a new method for measuring the electron 
 energy by observing the radiation intensity near resonance. This will be
 discussed in more detail in a later paper. This method could also be used
 for ultra high energy  muons \cite{Palm}.

\newpage
 
 {\bf Figure captions}
\vskip 1cm

 1.Sketch  of the electron quasi--energy spectrum  for
   $\xi^{2}/\delta^{2}\ll 1$.
\vspace{5mm} 
 
 2.Sketch of the electron quasi--energy spectrum for
   $a\xi^{2}/\delta^{2}\gg 1$.
 

\vskip 1cm
 
\begin{thebibliography}{99}
\bibitem{redm-65}
 Redmond, P.J.,~~J.Math.Phys.~{\bf 6}~(1965)~1163.
\bibitem{berg-83}
 Bergou, J., Ehlotzky, F.,~~Phys.Rev.~{\bf A27}~(1983)~2291.
\bibitem{berest-80}
 Berestetskii, V.B., Lifshitz, E.M., Pitaevskii, L.P.,~{\it Quantum
 Electrodynamics}, Pergamon (1982), (Nauka,~Moscow,~1980).
\bibitem{Milan}
 Milant'ev, V.P.,~~Uspekhi Fiz. Nauk,~{\bf 167} $N^o$ 1 (1997) ~3 and \\
Phys.Uspekhi, ~{\bf 40} $N^o$ 1 (1971) 1.
\bibitem{ole-71}
 Oleinik, V.P.,~~ZhETP. ~{\bf 61}~(1971)~27 and \\
Sov.Phys.JETP., ~{\bf 34} $N^o$ 1 (1972) 14.
\bibitem{gold-64}
 Goldman, I.I.,~~Izv.Ak.Nauk.Arm(Phys.), ~{\bf 17}~(1964) ~129.
\bibitem{land-89}
 Landau, L.D.,~Lifshitz, E.M., {\it Quantum Mechanics},.
 ~Nauka,~Moscow,~1989.
\bibitem{sok-83}
 Sokolov, A.A.,~Ternov, I.M. and Kilminster C.W., {\it Radiation from 
 Relativistic Electrons}, American Institute of Physics (translation series),
 1986, \\(Nauka,~Moscow,~1983).
\bibitem{zeld-66}
 Zeldovich, Ya.B.,~~ZhETP. ~{\bf 51} ~(1966)~1492 and \\
Sov.Phys.JETP., {\bf 24} $N^o$ 5 (1967) 1006.
\bibitem{nik-64}
 Nikishov, A.I.,~Ritus, B.I.,~~ZhETP. ~{\bf 46} ~(1964) ~776 and \\
Sov.Phys.JETP., {\bf 19} $N^o$ 2 (1964) 529.
\bibitem{conn-68}
 O'Connel, R.F.,~~Phys.Lett., ~{\bf 27A} ~(1968) ~391.
\bibitem{Bagrov}  
 Bagrov, V.G.,~Khalilov, V.R.,~~Izv.Vuz.Phys.,~${N^o}$ 2, (1968) ~37.
\bibitem{spin98}
Melikian, R.A.,~~Barber, D.P., in the proceedings of
the 13th International Symposium on
High Energy Spin Physics (SPIN98), Protvino, Russia, September 1998. 
\bibitem{Palm}
 Palmer, R., et al., CERN 96-07, (1996) ~887. 
\end{thebibliography}
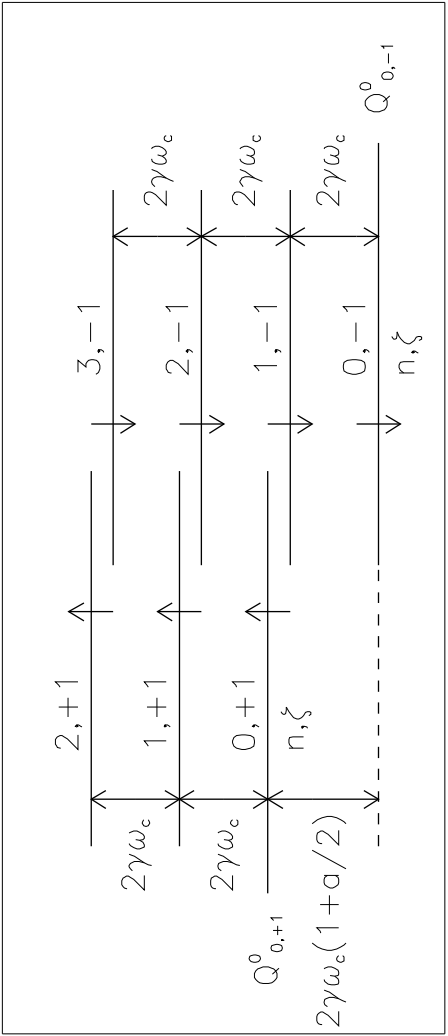
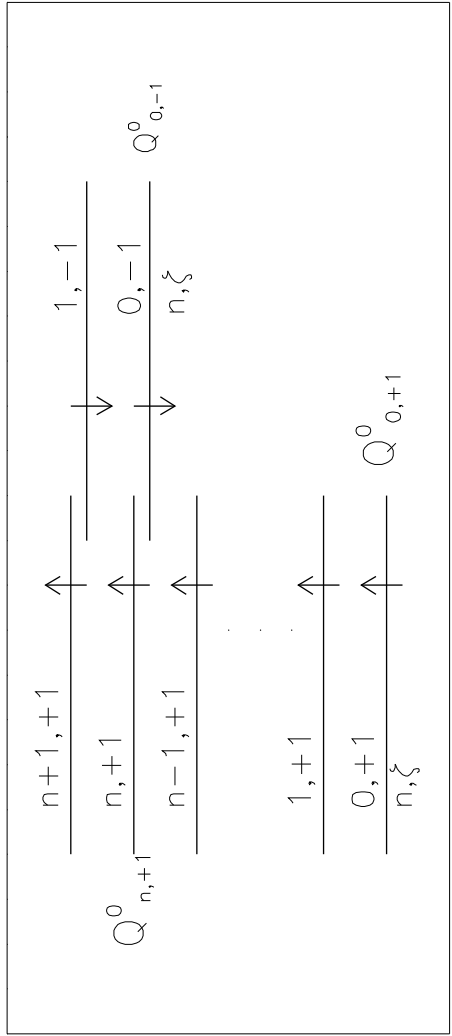
\end{document}